\documentclass[twocolumn,aps,pr,superscriptaddress,preprintnumbers,nofootinbib,10pt]{revtex4-2}
\usepackage{amsmath,amssymb}
\usepackage{amsmath,amssymb}
\usepackage{graphicx}
\usepackage{hyperref}
\usepackage{url}
\usepackage{slashed}
\usepackage{subcaption}
\usepackage[usenames,dvipsnames]{xcolor}
\usepackage{amsfonts}
\usepackage{float}
\usepackage{euscript}
\usepackage[utf8]{inputenc}
\allowdisplaybreaks
\usepackage[normalem]{ulem}
\usepackage{pifont}
\usepackage{dsfont}
\usepackage{verbatim}
\usepackage{latexsym}
\usepackage{courier}

\usepackage{tikz-feynman}
\usepackage[utf8]{inputenc}
\allowdisplaybreaks
\usepackage[normalem]{ulem}
\usepackage{pifont}
\usepackage{dsfont}
\usepackage{MnSymbol}
\usepackage{verbatim}
\usepackage{graphicx}
\usepackage{latexsym}
\usepackage{courier}

\usepackage{tikz-feynman}

\hypersetup{
colorlinks=true,
citecolor=blue,
citebordercolor=red,
linktoc=all,
linkcolor=blue,
urlcolor=blue
}

\def \s{\mathsf{s}}

\def \and{\textmd{and}}

\def \be{\begin{equation}}
\def \ee{\end{equation}}

\def \bea{\begin{eqnarray}}
\def \eea{\end{eqnarray}}

\newbox{\ORCIDicon}
\sbox{\ORCIDicon}{\large
                  \includegraphics[width=0.8em]{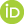}}

\begin{document}

\title{The Gribov problem on the horizon:\\ A short guide to the Gribov-Zwanziger framework and its refinement}

\author{Antonio D. Pereira\,\href{https://orcid.org/0000-0002-6952-2961}{\usebox{\ORCIDicon}}} \email{adpjunior@id.uff.br}
\affiliation{Instituto de F\'isica, Universidade Federal Fluminense, Campus da Praia Vermelha, Av. Litor\^anea s/n, 24210-346, Niter\'oi, RJ, Brazil}

\begin{abstract}
Gauge fixing in Yang–Mills theories encounters a non-perturbative obstruction: distinct gauge-equivalent configurations can satisfy the same gauge condition. These are the so-called Gribov copies and their existence challenges the global validity of the Faddeev–Popov prescription. Addressing this ambiguity motivates a framework that connects the geometry of gauge-field space to infrared dynamics. This mini-review introduces the Gribov problem in Euclidean Yang–Mills theories and traces the construction of the Gribov–Zwanziger action and its refinement through the introduction of dimension-two condensates. We discuss the resulting infrared propagators, their comparison to lattice results, and the role of BRST symmetry, before surveying selected applications and open questions.
\end{abstract}

\maketitle

\section{Introduction\label{Sec:Intro}}

The standard continuum quantization of Yang-Mills (YM) theories employs gauge fixing to handle gauge redundancy and define the gauge-field propagator. This is achieved through the Faddeev-Popov (FP) procedure \cite{Faddeev:1967fc}.  Remarkably, after fixing the gauge, a global symmetry emerges, the so-called BRST symmetry \cite{Becchi:1975nq,Tyutin:1975qk}. This symmetry allows for a systematic control of perturbative renormalizability, gauge dependence, as well as the characterization of the physical subspace of the theory, see \cite{Baulieu:1983tg}. Perturbative computations are successfully performed within this framework. In particular, four-dimensional pure YM theories are asymptotically free, a property that makes perturbation theory reliable at sufficiently high energies, \cite{Politzer:1973fx,Gross:1973id}. In the infrared, however, the theory becomes strongly coupled and ordinary perturbation theory becomes unrealible. Moreover, there is substantial evidence for confinement in pure YM, but a complete understanding of its dynamical origin remains a central open problem, see, e.g., \cite{Greensite:2011zz,Brambilla:2014jmp}.

Beyond perturbation theory, the gauge-fixing procedure itself encounters an obstruction. A gauge condition can admit distinct gauge-equivalent configurations, known as Gribov copies, so that the FP prescription does not completely remove gauge redundancy. This ambiguity is not merely a consequence of a poorly chosen gauge: The topology of the space of gauge orbits obstructs a globally valid, continuous choice of a unique representative per gauge orbit, \cite{Gribov:1977wm,Singer:1978dk}. These difficulties constitute the Gribov problem.

In this mini-review, we introduce the Gribov problem in Euclidean YM theories, focusing on the Gribov-Zwanziger framework and its refined formulation. We discuss their implications for infrared correlation functions and briefly survey selected applications and open questions.

\section{Faddeev–Popov quantization\label{Sec:EYM}}

The YM action in four-dimensional Euclidean space is\footnote{We employ the shorthand notation $\int {\rm d}^4x \equiv \int_x$. Likewise, $\int_q \equiv \int \frac{{\rm d}^4q}{(2\pi)^4}$ for integrals in momentum space.}
\begin{equation}
S_{\rm YM} = \frac{1}{4}\int_x~F^{a}_{\mu\nu} F^{a}_{\mu\nu}\,,
\label{Eq.EYM.1}
\end{equation}
with $F^a_{\mu\nu} = \partial_\mu A^a_\nu - \partial_\nu A^a_\mu + gf^{abc}A^b_\mu A^c_\nu$ being the field strength associated with the gauge field $A^a_\mu$. The coupling constant $g$ is dimensionless and $f^{abc}$ are the totally antisymmetric structure constants of $SU(N)$. Action \eqref{Eq.EYM.1} is invariant under the gauge transformation
\begin{equation}
A^\prime_\mu \equiv A^U_\mu = U A_\mu U^\dagger + \frac{i}{g} U \partial_\mu U^\dagger\,.
\label{Eq.EYM.2}
\end{equation}
In Eq.~\eqref{Eq.EYM.2}, $A_\mu \equiv A^a_\mu T^a$, with $\left\{T^1,\ldots,T^{N^2-1}\right\}$ being the generators of $SU(N)$. The local elements of $SU(N)$ are $U = {\rm exp }(-ig \alpha^a (x) T^a)$. For infinitesimal gauge transformations, Eq.~\eqref{Eq.EYM.2} reduces to
\begin{equation}
A^{\prime a}_\mu = A^a_\mu - D^{ab}_\mu \alpha^b\,,
\label{Eq.EYM.3}
\end{equation}
with $D^{ab}_\mu \equiv \delta^{ab}\partial_\mu - gf^{abc}A^c_\mu$ being the covariant derivative in the adjoint representation. Formally, the path integral is written as
\begin{equation}
\EuScript{Z} = \int [\EuScript{D}A]\,{\rm e}^{-S_{\rm YM}}\,,
\label{Eq.EYM.4}
\end{equation}
where all gauge-field configurations, including those that are connected by gauge transformations \eqref{Eq.EYM.2}, are integrated over. 

The FP procedure aims at rewriting the functional measure as a measure over inequivalent field configurations, i.e., over gauge fields that are not connected by gauge transformations. This requires selecting one representative per gauge orbit, where an orbit is the set of gauge-equivalent configurations. To this end, one imposes a gauge-fixing condition, $\mathcal{F}[A] = 0$, such as
\begin{equation}
\mathcal{F}^a[A] = \partial_\mu A^a_\mu\,.
\label{Eq.EYM.5}
\end{equation}
This is the Landau gauge condition, which we use as a concrete example throughout this section. The core of the FP procedure lies in the introduction of an identity in \eqref{Eq.EYM.4} which involves a delta-functional that enforces the condition \eqref{Eq.EYM.5} over the gauge fields. Formally, this is written as
\begin{equation}
1 = \Delta_{\rm FP} [A] \int [\EuScript{D}\mathfrak{g}]\,\delta (\mathcal{F}[A^\mathfrak{g}])\,,
\label{Eq.EYM.6}
\end{equation}
with $[\EuScript{D}\mathfrak{g}]$ being the invariant measure over the gauge group and $\Delta_{\rm FP} [A]$ is the determinant of the so-called FP operator. The assumptions underlying this identity can be illustrated by the ordinary-variable relation
\begin{equation}
1 = \left(\sum^{n}_{i=1} \frac{1}{|f^\prime (x_i)|}\right)^{-1} \int {\rm d}x\, \delta (f(x))\,,
\label{Eq.EYM.7}
\end{equation}
where $f(x_i) = 0$, $i=1,\ldots ,n$ and the derivative of $f(x)$ exists and does not vanish at each $x_i$. Two assumptions in the FP construction can be understood through this analogy. First, the gauge-fixing condition selects one and only one representative per gauge orbit. This is precisely the goal of gauge fixing. In \eqref{Eq.EYM.7} this assumption implies that the function $f(x)$ has only one zero, i.e., $n=1$. The second assumption is that $\Delta_{\rm FP}[A]$ is positive. It translates into assuming that the derivative of $f(x)$ computed at its zero (which we shall call $\bar{x}$) is positive. Hence, $|f^\prime (\bar{x})| = f^\prime (\bar{x})$. Under these assumptions, Eq.~\eqref{Eq.EYM.7} has the same structure as Eq.~\eqref{Eq.EYM.6}. In the Landau gauge, the FP operator $\mathcal{M}(A)$ is 
\begin{equation}
\mathcal{M}^{ab} = -\partial_\mu D^{ab}_\mu = -\delta^{ab}\partial^2 + gf^{abc}A^c_\mu \partial_\mu\,,
\label{Eq:EYM.9}
\end{equation}
where $\partial_\mu A^a_\mu = 0$. 

Following the standard steps of the FP procedure, one inserts \eqref{Eq.EYM.6} in \eqref{Eq.EYM.4}. The gauge-group integral factors out and the resulting path integral becomes
\begin{equation}
\EuScript{Z} = \mathcal{N}\int [\EuScript{D}A]\,\delta (\mathcal{F}[A])\,\Delta_{\rm FP} [A]\,{\rm e}^{-S_{\rm YM}}\,,
\label{Eq.EYM.10}
\end{equation}
with $\mathcal{N}$ being an overall normalization constant. The delta functional and the FP determinant can be exponentiated using the Nakanishi-Lautrup field $b^a$ which enforces the gauge condition and the Grassmann-valued FP ghosts $(\bar{c},c)^a$, whose integration reproduces the determinant factor. Thus, the gauge-fixed path integral can be written as
\begin{equation}
\EuScript{Z} = \int [\EuScript{D}\mu]_{\rm YM}\,{\rm e}^{-S_{\rm FP}}\,,
\label{Eq.EYM.11}
\end{equation}
with $[\EuScript{D}\mu]_{\rm YM} = [\EuScript{D}A][\EuScript{D}b][\EuScript{D}c][\EuScript{D}\bar{c}]$ and
\begin{equation}
S_{\rm FP} = S_{\rm YM} + \int_x \Big(ib^a \mathcal{F}^a [A] - \bar{c}^a \mathcal{M}^{ab}(A) c^b\Big)\,.
\label{Eq.EYM.12}
\end{equation}

Remarkably, the gauge-fixed action \eqref{Eq.EYM.12} is invariant under BRST transformations defined by
\begin{align}
\s A^a_\mu &= - D^{ab}_\mu c^b\,, && \s c^a = \frac{g}{2}f^{abc}c^b c^c\,, \nonumber\\
\s \bar{c}^a &= ib^a\,,  && \s b^a = 0\,,
\label{Eq.EYM.13}
\end{align}
with $\s$ being the nilpotent BRST operator, i.e., $\s^2 = 0$. The corresponding gauge-field propagator at tree level in the Landau gauge is
\begin{equation}
\langle A^a_\mu (p) A^{b}_\nu (-p) \rangle = \frac{\delta^{ab}}{p^2}\left(\delta_{\mu\nu} - \frac{p_\mu p_\nu}{p^2}\right)\,,
\label{Eq.EYM.14}
\end{equation}
which has the typical behavior of a massless and transverse vector field. 

In the next section, we show how Gribov copies invalidate the assumptions underlying the FP procedure beyond perturbation theory. 

\section{The Gribov Problem \label{Sec:GP}}

An ideal gauge condition selects exactly one representative per gauge orbit. To test this requirement in the Landau gauge, consider two configurations $A^a_\mu$ and ${A}^{\prime a}_\mu$ that are connected by an infinitesimal gauge transformation, see Eq.~\eqref{Eq.EYM.3}. If $A^a_\mu$ satisfies the Landau gauge condition, i.e., $\partial_\mu A^a_\mu = 0$, then a distinct configuration ${A}^{\prime a}_\mu$ on the same orbit should not satisfy it. From Eq.~\eqref{Eq.EYM.3}, the gauge field ${A}^{\prime a}_\mu$ satisfies the Landau gauge condition if
\begin{equation}
- \partial_\mu D^{ab}_\mu \alpha^b = 0\,,
\label{Eq.GP.2}
\end{equation}
i.e., if the FP operator has (normalizable) zero modes. In \cite{Gribov:1977wm}, Gribov showed that nontrivial zero modes exist establishing that a Landau-type gauge condition is not ideal. Field configurations that belong to the same gauge orbit and satisfy the same gauge condition are said to be Gribov copies and their existence establishes the Gribov problem. In fact, they also exist for gauge-field configurations that are connected by finite gauge transformations, see \cite{vanBaal:1991zw}. The Gribov problem is not a pathology of the Landau gauge but it is actually associated with the non-trivial bundle structure of the configuration space of non-Abelian gauge theories, see \cite{Singer:1978dk}. For comprehensive reviews on the Gribov problem we refer the reader to, e.g., \cite{Vandersickel:2012tz,Sobreiro:2005ec}. Effectively, the Gribov problem reveals that the FP procedure does not completely fix the gauge freedom. The existence of zero modes of the FP operator shows that both assumptions that are tacitly made in the FP method are not valid: normalizable zero modes indicate that more than one representative per gauge orbit satisfies the gauge-fixing condition and that the FP operator is singular. These difficulties do not invalidate the usual perturbative construction around $A_\mu^a=0$, where $\mathcal {M}\approx -\partial^2$, with trivial zero modes excluded. Away from this perturbative neighborhood, the field-dependent contribution to $\mathcal{M}$, see Eq.~\eqref{Eq:EYM.9}, can generate nontrivial zero modes. The success of perturbative calculations therefore does not establish the global validity of the FP prescription.

In \cite{Gribov:1977wm}, a method to deal with infinitesimal Gribov copies in the Landau gauge was proposed. It consists in restricting the gauge-fixed path integral to a region in field space where the FP operator is positive. In the Landau gauge the associated FP operator is Hermitian and, as such, it is meaningful to define such a region. It is known as the Gribov region and it is defined by
\begin{equation}
\Omega = \left\{A^a_\mu\, |\, \partial_\mu A^a_\mu = 0 \,\, {\rm and} \,\, -\partial_\mu D^{ab}_\mu > 0 \right\}\,.
\label{Eq.GP.3}
\end{equation}
The Gribov region has three important geometric properties: (a) It is bounded in every direction; (b) It is convex; (c) Every gauge orbit crosses it at least once, see \cite{DellAntonio:1991mms}. The boundary $\partial \Omega$ is known as the Gribov horizon and it is where the FP operator hits its first zero modes. The third property ensures that the restriction of the path integral to the Gribov region does not leave out any physically relevant information. However, it should be noted that the Gribov region is not free from all Gribov copies. In particular, there are those generated by finite gauge transformations, see \cite{vanBaal:1991zw}. Removing all copies requires restricting further the functional integral to the so-called Fundamental Modular Region (FMR). However, up to now, a practical implementation of the restriction to the FMR in the continuum does not exist. The proposal of \cite{Gribov:1977wm} consists in implementing the following,
\begin{equation}
\EuScript{Z} = \int_{\Omega} [\EuScript{D}\mu]_{\rm YM}\,{\rm e}^{-S_{\rm FP}}\,.
\label{Eq.GP.4}
\end{equation}

In the next section, we outline the original strategies used to implement the restriction \eqref{Eq.GP.4}.

\section{The No-pole condition and the Gribov-Zwanziger action \label{Sec:NPHF}}

Gribov’s restriction of the Landau-gauge functional integral relies on the relation between the FP operator and the ghost two-point function. Extracting the ghost propagator from \eqref{Eq.GP.4} leads to
\begin{eqnarray}
\langle \bar{c}^{a}(x)c^{b}(y)\rangle &=& \mathcal{N}\int \left[\EuScript{D}A\right]\EuScript{V}(\Omega)\;\delta(\partial_{\mu}A^{a}_{\mu})\mathrm{det}(\mathcal{M})\nonumber\\
&\times& \left[\mathcal{M}^{-1}\right]^{ab}(x,y)\,\mathrm{e}^{-S_{\mathrm{YM}}}\,,
\label{Eq.NPHF.1}
\end{eqnarray} 
where the ghosts and the Nakanishi-Lautrup field have been integrated out. The factor $\EuScript{V}(\Omega)$ is responsible for enforcing the restriction to $\Omega$. At fixed gauge field, the connected ghost propagator is
\begin{eqnarray}
\langle \bar{c}^{a}(x)c^{b}(y)\rangle_A &=& \left[\mathcal{M}^{-1}\right]^{ab}(x,y)\,.
\label{Eq.NPHF.2}
\end{eqnarray} 
In Fourier space in the limit $g A\to 0$, the ghost propagator form factor is $1/p^2$. Thus it is positive and the only pole is at $p^2=0$. Introducing the dependence on $gA$, one can define ghost form factor at fixed gauge-field configuration in momentum space as
\begin{eqnarray}
\mathcal{G}(p;A)  &\equiv& \frac{\langle \bar{c}^{a}(p)c^{a}(-p)\rangle_A}{N^2-1}\,,
\label{Eq.NPHF.3}
\end{eqnarray} 
which can be computed order by order in insertions of $gA$. Since the Gribov region is defined by the positivity of the FP operator and the first zero modes sit at its boundary, we can formulate the restriction of the path integral to $\Omega$ by imposing that the inverse FP operator, i.e., the ghost two-point function at fixed gauge-field configuration, develops no poles besides $p^2=0$. This is known as Gribov's no-pole condition and can be imposed on $\mathcal{G}(p;A)$. In \cite{Gribov:1977wm}, this was computed at leading order in an expansion in powers of $gA$. At leading nontrivial order in the background field, the ghost form factor can be written in the approximate resummed form
\begin{equation}
\mathcal{G}(p;A)\approx\frac{1}{p^2}\frac{1}{1-\sigma(p;A)}\,,
\label{Eq.NPHF.6}
\end{equation}
with
\begin{equation}
\sigma(p;A)=\frac{1}{V}\frac{1}{p^2}\frac{Ng^2}{N^2-1}\int_q A^{a}_{\mu}(-q)A^{a}_{\nu}(q)\frac{(p-q)_{\mu}p_{\nu}}{(p-q)^2}\,,
\label{Eq.NPHF.5}
\end{equation}
where $V$ denotes the spacetime volume. See  \cite{Sobreiro:2005ec,Vandersickel:2012tz} for details. Following the strategy of \cite{Gribov:1977wm}, $\sigma(p;A)$ captures the pole structure of \eqref{Eq.NPHF.6}.
In fact, within the present approximation, imposing that \eqref{Eq.NPHF.6} does not develop poles besides $p^2 = 0$ amounts to enforce $\sigma (0;A) < 1$, where
\begin{equation}
\sigma(0,A)=\frac{1}{4V}\frac{Ng^2}{N^2-1}\int_q \frac{A^{a}_{\lambda}(-q)A^{a}_{\lambda}(q)}{q^{2}}\,.
\label{Eq.NPHF.10}
\end{equation}
This requirement is the so-called no-pole condition, see \cite{Gribov:1977wm,Sobreiro:2005ec,Vandersickel:2012tz}. Thus, a concrete expression for $\EuScript{V}(\Omega)$ is $\EuScript{V}(\Omega)=\theta(1-\sigma(0;A))$ with $\theta (x)$ being the Heaviside step function. It implements the condition $\sigma (0;A) < 1$. The step function can be exponentiated using its integral representation, i.e.,
\begin{equation}
\EuScript{V}(\Omega)=\theta(1-\sigma(0,A))=\int^{+i\infty+\epsilon}_{-i\infty+\epsilon}\frac{{\rm d}\beta}{2\pi i\beta}\mathrm{e}^{\beta (1-\sigma(0,A))}\,.
\label{Eq.NPHF.9}
\end{equation}
Since $\sigma(0;A)$ is quadratic in the gauge field, it contributes, for instance, to the gauge-field propagator. After retaining quadratic terms and integrating over the gauge field, the remaining $\beta$ integral is evaluated by steepest descent, see \cite{Gribov:1977wm,Sobreiro:2005ec,Vandersickel:2012tz}. Its saddle point $\beta_0$ satisfies
\begin{equation}
\frac{1}{\beta_0}+\frac{3Ng^2}{4}\int_q \frac{1}{q^4 + \frac{\beta_0 N g^2}{2V(N^2-1)}} = 1\,.
\label{Eq.NPHF.13}
\end{equation}
We define the Gribov parameter $\lambda$ by $\lambda^4 \equiv {\beta_0 N g^2}/{2V(N^2-1)}$. In order to have a finite value for $\lambda$, the parameter $\beta_0$ should scale as $\beta_0 \sim V$. Hence, the term $1/\beta_0$ is suppressed with respect to the integral in \eqref{Eq.NPHF.13} and can be neglected in the thermodynamic limit. Thus, Eq.~\eqref{Eq.NPHF.13} reduces to the gap equation
\begin{equation}
\frac{3Ng^2}{4}\int_q \frac{1}{q^4+\lambda^4} = 1\,.
\label{Eq.NPHF.15}
\end{equation}
After regularization and renormalization, the gap equation determines the Gribov parameter in terms of the renormalized coupling and renormalization scale. One can plug the solution of \eqref{Eq.NPHF.15} back into the path integral restricted to the Gribov region. In this case, the tree-level gauge-field propagator reads
\begin{equation}
\langle A^a_\mu (p) A^b_\nu (-p)\rangle = \delta^{ab}\frac{p^2}{p^4 + \lambda^4}\left(\delta_{\mu\nu}-\frac{p_\mu p_\nu}{p^2}\right)\,,
\label{Eq.NPHF.16}
\end{equation}
The Gribov parameter acts as an infrared regulator to the propagator. In particular, at vanishing momentum, the gauge-field propagator vanishes. Additionally, the presence of the Gribov parameter generates a pair of complex-conjugate poles $p^2=\pm i\lambda^2$. The last two properties imply violation of reflection positivity. Thus, the gauge-field excitations cannot be interpreted as stable particles, a property often interpreted as a manifestation of confinement, see \cite{Zwanziger:1991gz}.

At tree-level, the ghost propagator retains its FP form. Computing the leading correction and considering the limit of small momentum, the ghost propagator reads
\begin{equation}
\langle \bar{c}^a (p) c^b (-p)\rangle \stackrel{p\to 0}{\propto} \lambda^2\frac{\delta^{ab}}{p^4}\,.
\label{Eq.NPHF.17}
\end{equation}
The restriction to the Gribov region enhances the ghost propagator in the deep infrared, i.e., it goes as $1/p^4$ instead of $1/p^2$. The behavior of the gauge-field and ghost propagators just reported shares the qualitative behavior of the scaling-type solution, see, e.g., \cite{vonSmekal:1997ohs,Alkofer:2000wg}.

In \cite{Zwanziger:1988jt,Zwanziger:1989mf}, the restriction of the path integral was performed at all orders in perturbation theory by means of a different strategy.  Zwanziger analyzed the spectrum of the FP operator directly and implemented the restriction to $\Omega$ by enforcing the positivity of its lowest eigenvalues. This condition is enforced by introducing in the Boltzmann weight the non-local horizon function $H(A)$ 
\begin{equation}
H(A) = g^2 \int_{x,y} f^{abc}A^b_\mu (x) \left[\mathcal{M}^{-1}\right]^{ad} (x,y) f^{dec}A^e_\mu (y)\,.
\label{Eq.NPHF.21}
\end{equation}
The partition function becomes
\begin{equation}
\EuScript{Z} = \int[\EuScript{D}\mu]_{\rm YM}\,{\rm e}^{-S_{\rm FP}-\gamma^4 H(A)+4V\gamma^4(N^2-1)}\,.
\label{Eq.NPHF.22}
\end{equation}
The parameter $\gamma$ relates the Gribov parameter $\lambda$ by $\lambda^4 = 2Ng^2\gamma^4$. It is not free but determined by the so-called horizon condition,
\begin{equation}
\langle H(A)\rangle = 4V (N^2-1)\,.
\label{Eq.NPHF.23}
\end{equation}
The detailed derivation of \eqref{Eq.NPHF.22} can be found in \cite{Vandersickel:2012tz}. The exponent in Eq.~\eqref{Eq.NPHF.22} together with condition \eqref{Eq.NPHF.23} effectively restricts the path integral to the Gribov region. Although the restriction to the Gribov region using the no-pole condition follows a different strategy, it is possible to prove that the extension of the no-pole condition to all orders leads to the same result as \eqref{Eq.NPHF.22}, see \cite{Capri:2012wx}.

The horizon function in \eqref{Eq.NPHF.21} is non-local. Yet it is possible to localize it by means of auxiliary fields, see \cite{Zwanziger:1989mf}. Thus, the gauge-fixed path integral restricted to the Gribov region reads
\begin{equation}
\EuScript{Z} = \int [\EuScript{D}\mu]_{\rm GZ}\,{\rm e}^{-S_{\rm GZ} + 4V\gamma^4 (N^2-1)}\,,
\label{Eq.NPHF.24}
\end{equation}
with
\begin{eqnarray}
S_{\rm GZ} &=& S_{\rm FP} - \int_x \left(\bar{\varphi}^{ac}_\mu \mathcal{M}^{ab}\varphi^{bc}_\mu - \bar{\omega}^{ac}_\mu \mathcal{M}^{ab} \omega^{bc}_\mu\right)\nonumber\\
&-&g\int_x f^{abc}\partial_\mu\bar{\omega}^{ae}_\nu (D^{bd}_\mu c^d) \varphi^{ce}_\nu 
\nonumber\\
&+& g\gamma^2 \int_x f^{abc}A^a_\mu (\varphi + \bar{\varphi})^{bc}_\mu\,.
\label{Eq.NPHF.25}
\end{eqnarray}
The fields $(\bar{\varphi},\varphi)$ are commuting and complex and $(\bar{\omega},\omega)$ are Grassmann-valued. The measure $[\EuScript{D}\mu]_{\rm GZ}$ is defined as $[\EuScript{D}\mu]_{\rm GZ} = [\EuScript{D}\mu]_{\rm YM} [\EuScript{D}\varphi] [\EuScript{D}\bar{\varphi}][\EuScript{D}\omega][\EuScript{D}\bar{\omega}]$. The action $S_{\rm GZ}$ is known as the Gribov-Zwanziger (GZ) action and provides a local and renormalizable framework that dynamically implements the restriction of the functional measure to the Gribov region, see \cite{Zwanziger:1989mf,Vandersickel:2012tz}. Within this local framework, the horizon condition becomes $\partial \EuScript{E}_v/\partial \gamma^2 = 0$, 
with $\EuScript{E}_v$ denoting the vacuum energy. This equation fixes the Gribov parameter $\gamma$. The auxiliary localizing fields are introduced as BRST doublets, i.e., 
\begin{align}
\s \bar{\omega}^{ab}_\mu & = \bar{\varphi}^{ab}_\mu\,, && \s \bar{\varphi}^{ab}_\mu = 0\,, \nonumber\\
\s \varphi^{ab}_\mu &= \omega^{ab}_\mu\,,  && \s  \omega^{ab}_\mu = 0\,.
\label{Eq.NPHF.27}
\end{align}
The transformations \eqref{Eq.NPHF.27} together with \eqref{Eq.EYM.13} do not lead to a symmetry of $S_{\rm GZ}$, i.e., $\s S_{\rm GZ} \neq 0$. In other words, the GZ action breaks BRST symmetry. Such a breaking has the form
\begin{equation}
\s S_{\rm GZ} = g\gamma^2 \int_x f^{abc}\left[A^a_\mu \omega^{bc}_\mu - D^{ad}_\mu c^d (\varphi + \bar{\varphi})^{bc}_\mu\right]\,.
\label{Eq.NPHF.28}
\end{equation}
Albeit explicit, the breaking \eqref{Eq.NPHF.28} is soft since it is proportional to $\gamma^2$. In the deep ultraviolet, the Gribov parameter is suppressed and BRST invariance is effectively restored. The fate of BRST symmetry in the strongly-coupled regime and the corresponding consequences have been thoroughly investigated, see, e.g., \cite{Maggiore:1993wq,Dudal:2009xh,Sorella:2009vt,Schaden:2014bea}. 

In \cite{Capri:2015ixa}, the GZ action was reformulated in terms of a non-local, gauge-invariant and transverse dressed field $A^h_\mu$. Within this framework, BRST invariance is restored. This has enabled the construction of a local and BRST-invariant formulation of the GZ action in the Landau gauge which is extendible to a large class of covariant gauges, see, e.g., \cite{Pereira:2016fpn,Mintz:2018hhx}. As a consequence, the dressed version of the horizon function \eqref{Eq.NPHF.21} acquires some degree of universality, see \cite{Capri:2018ijg}. Remarkably, the local version of the BRST-invariant GZ action in linear covariant gauges is not polynomial in the elementary fields, but it is renormalizable, see \cite{Capri:2017bfd}. Thus, although its original formulation breaks BRST invariance explicitly, a reformulation which is compatible with BRST symmetry is available. In this formulation, the Gribov parameter does not couple to a BRST-exact term and therefore is not akin to a gauge-fixing parameter. Thus, it can affect BRST-invariant correlation functions, see \cite{Capri:2016aqq}. Additionally, this paves the way for the definition of the underlying physical subspace and control of gauge(-parameter) dependence.

In the next section we take into account further non-perturbative effects that lead to the so-called Refined Gribov-Zwanziger action.

\section{The Refined Gribov-Zwanziger framework \label{Sec:RGZ}}

The GZ framework predicts that the gauge-field propagator is infrared suppressed and vanishes at zero momentum in the Landau gauge. As for the ghost propagator, it is enhanced at low energies. However, gauge-fixed lattice simulations revealed a different behavior for such correlation functions, see, e.g., \cite{Cucchieri:2007rg,Cucchieri:2008fc}. The gauge-field propagator remains infrared suppressed but approaches a non-vanishing value at zero momentum and the ghost propagator is not enhanced at low energies. Such propagators are said to be of massive- or decoupling-type. In fact, besides the scaling solution, functional methods also reported on the decoupling solution, see, e.g., \cite{Aguilar:2008xm,Fischer:2008uz} and \cite{Huber:2018ned}. It is natural to ask if the GZ framework is able to describe such an infrared behavior of the gauge-field and ghost propagators. 

Gribov copies potentially play an important role at low energies. However, there are other non-perturbative effects that may also need to be included in the underlying framework. One particular example is the dynamical formation of condensates. In the context of gauge-fixed pure YM theories the condensate $\langle A^2\rangle$ has been extensively investigated in the Landau gauge with supporting evidence for its formation, see \cite{Verschelde:2001ia,Boucaud:2001st}. The localizing auxiliary fields also participate in the dynamics and can form condensates as well. Effective-potential analyses indicate that the GZ vacuum favors the formation of dimension-two condensates. This was proposed and investigated in \cite{Dudal:2007cw,Dudal:2008sp}. The resulting action that takes into account the elimination of infinitesimal Gribov copies and the dynamical formation of condensates in the Landau gauge is the so-called Refined GZ (RGZ) action and is given by
\begin{equation}
S_{\rm RGZ} = S_{\rm GZ} + \frac{m^2}{2}\int_x A^a_\mu A^a_\mu - M^2 \int_x (\bar{\varphi}^{ab}_\mu \varphi^{ab}_\mu - \bar{\omega}^{ab}_\mu \omega^{ab}_\mu)\,,
\label{Eq.RGZ.1}
\end{equation}
where $m^2$ and $M^2$ are the mass parameters associated with the condensates $\langle A^2\rangle$ and $\langle \bar{\varphi}\varphi - \bar{\omega}\omega\rangle$. In a self-consistent treatment, these parameters are determined together with $\gamma$ by the minimization of the effective potential. Yet in phenomenological applications, they are usually extracted from lattice fits. They are not free but dynamically fixed by their own gap equations, see, e.g., \cite{Dudal:2011gd,Dudal:2019ing}. The tree-level gauge-field propagator obtained from \eqref{Eq.RGZ.1} is
\begin{equation}
\langle A^a_\mu (p) A^b_\nu (-p)\rangle = \delta^{ab}\mathcal{D}(p^2)\left(\delta_{\mu\nu}-\frac{p_\mu p_\nu}{p^2}\right)\,,
\label{Eq.RGZ.2}
\end{equation}
with
\begin{equation}
\mathcal{D}(p^2) = \frac{p^2+M^2}{(p^2+m^2)(p^2+M^2)+2Ng^2\gamma^4}\,.
\label{Eq.RGZ.3}
\end{equation}
Clearly, in the $p^2\to 0$ limit, the gauge-field propagator does not vanish as long as the mass-parameter $M^2$ associated with the condensate of auxiliary fields is non-zero. Evidence for this condensation already appears in a one-loop analysis around the GZ vacuum, see \cite{Dudal:2008sp}. One can show that the ghost propagator is not enhanced in the infrared. Thus, the RGZ action supports a decoupling-type solution for the gauge-field and ghost propagators in agreement with lattice data, see, e.g., \cite{Dudal:2010tf,Cucchieri:2011ig}. Additionally, for the parameter values relevant to lattice comparisons, the propagator violates reflection positivity. Evidence for the dynamical formation of such condensates is also observed in $d=3$ but not in $d=2$ due to infrared singularities, see, \cite{Dudal:2008rm,Dudal:2008xd}. Lattice simulations report a scaling-type propagator in $d=2$ in agreement with the standard GZ setup, see \cite{Cucchieri:2011ig}. As in the GZ scenario, the RGZ action \eqref{Eq.RGZ.1} breaks BRST invariance explicitly but it admits a BRST-invariant reformulation in terms of a dressed gauge-invariant field $A^h_\mu$, see, e.g., \cite{Capri:2015nzw}.

The RGZ framework was successfully employed in the estimation of glueball masses. Capturing non-perturbative information from fits of the mass parameters $(\gamma,m,M)$ from lattice data, the RGZ framework provides good estimates for low-lying glueball masses, see, e.g., \cite{Dudal:2010cd}. It has also been used in the computation of the Casimir energy in the MIT bag model leading to an expected negative value, see \cite{Canfora:2013zna}. See also \cite{Dudal:2026sue}. First steps on the interplay between the RGZ mass parameters and confinement/deconfinement transition were reported in \cite{Canfora:2015yia}. See \cite{Dudal:2023nbt} for recent related developments. Estimates for the topological susceptibility were worked out in \cite{Dudal:2017kxb}. One-loop calculations of the gluon propagator and ghost-gluon vertex using the interaction vertices of the RGZ action provide further quantitative comparisons with lattice data \cite{Mintz:2017qri,Barrios:2024idr,deBrito:2024ffa}. Extensions to matter propagators have also been investigated \cite{deBrito:2023qfs,deBrito:2025nvl}.

\section{Concluding Remarks and Perspectives\label{Sec:PerspConcl}}

The standard FP prescription does not completely fix the gauge beyond perturbation theory. Although a complete treatment of this residual redundancy remains challenging, substantial progress has been achieved. In the Landau gauge, restricting the functional integral to the Gribov region excludes infinitesimal copies and introduces a dynamically determined mass scale. The RGZ framework combines this restriction with dimension-two condensates in a local and renormalizable formulation, reproducing the qualitative infrared behavior of gluon and ghost propagators observed on the lattice. Non-perturbative BRST-invariant formulations have also enabled extensions beyond the Landau gauge. 

Important open questions include the role of Gribov copies remaining inside the Gribov region, the consequences of the restriction for gauge-invariant observables \cite{Mena:2024qou}, and the construction of a consistent Minkowski-space formulation of the (R)GZ framework \cite{Guimaraes:2026zsv}. The Gribov problem in (quantum) gravity also warrants further investigation \cite{Das:1978gk}.
 
\section*{Acknowledgments} 

The author acknowledges support from CNPq under the grant PQ-B (307623/2026-2) and from FAPERJ through the ``Jovem Cientista do Nosso Estado" program (E-26/204.457/2025)





\bibliography{refs}

\end{document}